\documentclass[sigconf]{acmart}

\renewcommand\footnotetextcopyrightpermission[1]{}

\setcopyright{acmlicensed}
\acmYear{2024}\copyrightyear{2024}
\acmConference[ASIA CCS '24]{ACM Asia Conference on Computer and Communications Security}{July 1--5, 2024}{Singapore, Singapore}
\acmBooktitle{ACM Asia Conference on Computer and Communications Security (ASIA CCS '24), July 1--5, 2024, Singapore, Singapore}
\acmDOI{10.1145/3634737.3637636}
\acmISBN{979-8-4007-0482-6/24/07}

\usepackage[T1]{fontenc}
\PassOptionsToPackage{hyphens}{url}\usepackage{hyperref}
\usepackage{cleveref}

\usepackage[english]{babel}
\usepackage{graphicx}
 \usepackage{booktabs}
\usepackage{xspace}
\usepackage[utf8]{inputenc}
\usepackage{wasysym}
\usepackage[ruled,vlined]{algorithm2e}
\usepackage{pdflscape}
\usepackage{enumitem}
\usepackage{verbatim}

\usepackage{subcaption}
\usepackage{float}
\usepackage{xcolor}
\usepackage{color}
\usepackage{colortbl}

\newcommand{\textbox}[1]{
    \noindent\fbox{%
        \parbox{0.98\columnwidth}{%
            #1
        }%
    }
}

\def \paragraph [#1] {\vspace{3pt}\noindent{\textbf{#1}\quad}}%

\newcommand{\pseudoparagraph}[1]{\vspace{3pt}\noindent\textbf{#1} --- }

\newcommand{\summary}[2]{
\vspace{0.5mm}
\textbox{
{\textbf{{#1}} {#2}}
}
}

\begin{document}

\title[What is in the Chrome Web Store? Investigating Security-Noteworthy Browser Extensions]
{What is in the Chrome Web Store?\\Investigating Security-Noteworthy Browser Extensions}

\author{Sheryl Hsu}
\affiliation{%
  \institution{Stanford University}
    \city{}
  \country{}
}
\email{sherylh@stanford.edu}

\author{Manda Tran}
\affiliation{%
  \institution{Stanford University}
    \city{}
  \country{}
}

\author{Aurore Fass}
\affiliation{%
  \institution{Stanford University, CISPA Helmholtz Center for Information Security}
    \city{}
  \country{}
}
\email{fass@cispa.de}

\renewcommand{\shortauthors}{Hsu et al.}

\begin{abstract}
This paper is the first attempt at providing a holistic view of the Chrome Web Store (CWS).
We leverage historical data provided by ChromeStats to study global trends in the CWS and security implications.
We first highlight the extremely \textit{short life cycles of extensions}: roughly 60\% of extensions stay in the CWS for one year.
Second, we define and show that \textit{Security-Noteworthy Extensions (SNE)} are a significant issue: they pervade the CWS for years and affect almost 350 million users.
Third, we identify \textit{clusters of extensions with a similar code base}. We discuss how code similarity techniques could be used to flag suspicious extensions. By developing an approach to extract URLs from extensions' comments, we show that extensions reuse code snippets from public repositories or forums, leading to the propagation of dated code and vulnerabilities.
Finally, we underline a critical \textit{lack of maintenance in the CWS}: 60\% of the extensions in the CWS have never been updated; half of the extensions known to be vulnerable are still in the CWS and still vulnerable 2 years after disclosure; a third of extensions use vulnerable library versions.
We believe that these issues should be widely known in order to pave the way for a more secure CWS.
\end{abstract}



\settopmatter{printfolios=true}

\maketitle

\vspace{1cm}
\section{Introduction} \label{sec:intro}

Browser extensions provide additional functionality and customization for browsers. The most popular desktop browser Chrome (with a market share of 66\%~\cite{chrome-usage}) has almost 125k extensions, totaling over 1.6 billion active users~\cite{chromestats}.
Unfortunately, due to their often specialized or privileged capabilities, browser extensions can either be a tool or a target for attackers.
Specifically, attackers are developing \textit{malicious extensions} to, e.g., spread malware via malvertising~\cite{Xing2015}, track users~\cite{Weissbacher2017}, spy on users~\cite{Aggarwal2018}, or steal credentials and other sensitive information~\cite{Chen2018}.
At the same time, other extensions have inherently benign functionalities but \textit{contain vulnerabilities} which, if exploited, lead to, e.g., universal cross-site scripting or leaking of sensitive user data~\cite{fass2021doublex, Some2019, Yu2023}.
Finally, merely using extensions (independent of whether they have any known flaws) represents a \textit{privacy risk} for Web users. For example, it is possible to infer the set of extensions a user has installed, by observing side effects some extensions induce (browser extension fingerprinting)~\cite{Starov2017, Starov2019, Solomos2022-2, Laperdrix2021, Sjosten2017, SanchezRola2017, Solomos2022}.
This enables an attacker to track users across websites or infer sensitive information about them~\cite{Karami2020}.

To mitigate these issues, browser vendors review extensions prior to publication, e.g., Google engineers vet extensions before their deployment in the Chrome Web Store (CWS)~\cite{chrome-vetting}.
Despite this vetting process and a decade of work on securing extensions, malicious, vulnerable, and fingerprintable extensions are still found in the CWS~\cite{fass2021doublex,malicious-ex}. We argue that a potential solution to this issue requires a comprehensive understanding of the underlying ecosystem of browser extensions. Indeed, and perhaps surprisingly, very little is currently known about what lies in browser extension~galleries.

This paper provides a holistic view of the browser extension landscape within the CWS.
We focus on the CWS because Chrome is the most popular browser, and Chrome extensions are built using the WebExtensions API--a cross-browser technology compatible with Firefox, Opera, Microsoft Edge, etc~\cite{webapi}.

We begin our analyses by investigating overall trends in the CWS, along with potential security problems that could arise from having such a big and diverse code base.
We first highlight unexpected volatility in terms of extensions in the CWS, e.g., only 60\% of extensions are available for one year.
Second, we define and investigate ``Security-Noteworthy Extensions'' (SNE): we analyze malware-containing, policy-violating, and vulnerable extensions.
We find that these SNE are a significant problem: over 346 million users installed a SNE in the last 3 years (280M malware, 63M policy violation, and 3M vulnerable). In addition, these extensions are staying in the CWS \textit{for years}, making thorough vetting of extensions and notification of impacted users all the more critical.
Third, we uncover thousands of clusters of similar extensions, and we show that investigating extension codes for similarities may enable us to identify unknown SNE.
Subsequently, we develop an approach to extract URLs from extensions' comments and attribute instances of code reuse to these cited URLs. We show concrete evidence that extensions reuse code from public repositories or forums directly, and we exemplify how code reuse leads to the propagation of vulnerabilities.
Fourth, we show that extensions are globally not maintained: 60\% of the extensions in the CWS have never been updated. This has direct security consequences, e.g., half of the vulnerable extensions discovered in 2021~\cite{fass2021doublex} are still in the CWS and still vulnerable in 2023.
Equally worrisome is the fact that developers continue to use deprecated tools, even when (more) secure alternatives are available. For example, a third of extensions use JavaScript libraries with known vulnerabilities, impacting almost 500 million users.

To sum up, our paper makes the following contributions:
\begin{itemize}[leftmargin=*]
\item We analyze overall trends in the CWS and highlight the exceptionally \textit{short life cycles of extensions};
\item We define and investigate \textit{``Security-Noteworthy Extensions`` (SNE)}. We show that these are a significant issue, affecting hundreds of millions of users and staying in the CWS for years;
\item We discover \textit{clusters of extensions with a similar code base} and highlight security implications;
\item We characterize a critical \textit{lack of maintenance in the CWS} and discuss the pervasiveness of vulnerable extensions in the CWS.
\end{itemize}

We are confident that our findings will guide future research and pave the way for a more secure CWS.

\section{Background: Browser Extensions} \label{sec:background}

Browser extensions are third-party programs that users can install to extend and customize their browser functionality by, e.g., adding ad-blocking capabilities or checking grammar.
In this section, we first give an overview of extension architecture and permission system. Then, we describe the main components of extensions.

\pseudoparagraph{Overview}
Browser extensions are zipped bundles of, e.g., HTML, JavaScript, or CSS files, stored in CRX files.
Every extension requires a JSON-formatted file, named \texttt{manifest.json}~\cite{manifest}, which specifies important information such as the extension's most important components and the extension's permissions. In fact, to use most Chrome APIs, e.g., \texttt{activeTab}, \texttt{downloads}, or \texttt{storage}, an extension must declare the corresponding permissions in its manifest~\cite{permissions}.
There are three versions of the manifest~\cite{manifest-versions}. Manifest V1 has been deprecated since 2012. Manifest V2, while deprecated, is still used by almost 62\% of extensions as of July 2023\cite{chromestats-v3}. It is currently unclear when Chrome will stop providing support for this version, as the deadline has been extended several times~\cite{manifest-v2}. Manifest V3 was released in November 2020 to improve the security, privacy, and performance of Chrome extensions~\cite{manifestv3-info}.
For example, manifest V3 prevents extensions from downloading external resources; instead, all resources must be bundled within the extension package.

\pseudoparagraph{Main Components}
The source code of browser extensions is split into several components, which are specified in the extension manifest file. The core logic of an extension is implemented through \textit{service workers} (\textit{background scripts} for manifest V2 extensions), which run independently of the lifetime of a web page and do not need any user interactions~\cite{serviceworker}.
An extension can inject \textit{content scripts} to run in the context of web pages, similarly to the scripts web pages directly load. While content scripts can use standard DOM APIs to read and modify web pages, they live in an ``isolated world'' to avoid conflicting with variables defined in web pages~\cite{cs}.
An extension can propose \textit{UI} or \textit{option pages} to enable users to customize their extension's behavior~\cite{uip}.
Finally, an extension can expose \textit{Web Accessible Resources (WARs)}, i.e., files that can be accessed by web pages or other extensions.

\section{Extension Collection and Analysis}

To study global trends in the CWS and corresponding security implications, we first need to collect a comprehensive set of extensions.
Since Google does not archive browser extensions that used to be in the CWS but are now deleted, we use ChromeStats~\cite{chromestats} to collect historical data.
We focus on likely-benign (i.e., not currently known to have any security or privacy issues) and security-noteworthy extensions (SNE): malware-containing, privacy-violating, and vulnerable extensions.
In this section, we first discuss the collection of our datasets and then analyze overall trends in the CWS. Finally, we dissect the life cycles of extensions in the CWS.

\subsection{Extension Collection} \label{subsec:ext-collection}

\pseudoparagraph{ChromeStats} \label{subsubsec:chromestats-dataset}
ChromeStats~\cite{chromestats} provides historical data since July~5, 2020 for browser extensions that are or were in the CWS.
ChromeStats developers designed a crawler to automatically collect extensions and extract their metadata from the CWS once a day. Besides the source code, they also archive, e.g., extension id, name, category, last update, number of users, or permission information.
To conduct longitudinal experiments and study global trends in the CWS, we collected all extensions from the CWS between July~5, 2020 and February 14, 2023. In this setting, we do not discriminate between benign and security-noteworthy extensions but consider all available extensions, independently of their intent.

\pseudoparagraph{Security-Noteworthy Extensions (SNE)} \label{subsec:sne}
The CWS contains what we call ``security-noteworthy extensions'' (SNE).
We define SNE as extensions known to either:

\begin{itemize}[leftmargin=*]
    \item \textbf{contain malware}: such extensions aim to, e.g., steal user-sensitive data, track users, spy on them, or propagate malware~\cite{Chen2018, Weissbacher2017, Aggarwal2018};
    \item \textbf{violate the CWS policies}: all extensions submitted to the CWS have to comply with the developer program policies~\cite{review-basics, program-policies};
    \item \textbf{contain vulnerabilities}: if exploited, such vulnerabilities could lead to, e.g., universal XSS or user data exfiltration~\cite{fass2021doublex, Some2019, Yu2023}.
\end{itemize}

\begin{table}
\begin{center}
\resizebox{\columnwidth}{!}{%
\footnotesize
\begin{tabular}{l r r l}
\toprule
Category & \begin{tabular}[c]{@{}l@{}}\# extensions\\ -- metadata collected\end{tabular} & \begin{tabular}[c]{@{}l@{}}\# extensions\\ -- code collected\end{tabular} & When collected \\
\midrule
SNE & 26,014 & 16,377 & \\
- Malware containing & 10,426 & 6,587 & July 5, 2020 -- May 1, 2023 \\
- Privacy violating & 15,404 & 9,638 & July 5, 2020 -- May 1, 2023 \\
- Vulnerable & 184 & 152 & March 16, 2021 \\
\midrule
Benign extensions & 226,762 & 92,482 & July 5, 2020 -- May 1, 2023 \\
\bottomrule
\end{tabular}
}
\end{center}
\caption{SNE and likely-benign extensions collected}
\label{tab:dataset-sne}
\end{table}

\noindent From ChromeStats, we extract a list of extensions that were removed from the CWS for containing malware or violating policies (such extensions were analyzed and flagged accordingly by Google engineers). 
We collected malware-containing and privacy-violating extensions that were removed from the CWS between July~5, 2020 and May 1, 2023.
To retrieve known vulnerable extensions, we contacted Fass et al.~\cite{fass2021doublex} who designed an advanced data flow analysis technique to automatically uncover vulnerabilities originating from PostMessages exchanged between web pages and browser extensions. We got access to their dataset of 184 vulnerable extensions (including the corresponding versions), which they collected on March 16, 2021.
To have a point of comparison, we also take the set of all 226,762 benign extensions that appeared in the CWS between July 5, 2020 and May 1, 2023. In this paper, we use the term \textit{benign} extensions to refer to extensions that are not known to be SNE.
As summarized in \Cref{tab:dataset-sne} (columns 1, 2, and 4), we retrieved metadata for 10,426 malware-containing extensions, 15,404 policy-violating, 184 vulnerable, and 226,762 benign extensions.

\pseudoparagraph{Extension Unpacking}
We used the ChromeStats API to download the source code of extensions, in the format of CRX files. As highlighted in \Cref{tab:dataset-sne} (column 3), we could download the source code of 16,377 out of 26,014 SNE (the rest was not available for download on ChromeStats). Given the large size of the benign extension set, we chose to only download benign extensions from 2023.
After downloading the extensions, we unpacked the corresponding CRX files~\cite{unpacker} to extract content scripts and service workers\,/\,background scripts. We chose to compile all content scripts and all background scripts into a single content and background file, respectively, similarly to prior work~\cite{fass2021doublex}.

\subsection{Overall Trends} \label{subsec:trends}

As of February 2023, there are 124,094 extensions in the CWS. We first discuss what users install extensions for, as well as the evolution of the number of active users and extensions over time.

\pseudoparagraph{What do Users Install Extensions for?}
As a proxy to investigate what users install extensions for, we consider the categories of the most popular extensions.
In the CWS, extensions are organized into categories which users can search through.
The most popular category is ``Productivity'' with a share of 41\%.
``Productivity'' contains a wide variety of extensions from translation tools to personal dashboards to PDF generators.
Of the 10 extensions with the highest number of active users (over 10M; as of July 2023), 9 are from the ``Productivity'' category; among these, 5 are content blockers (e.g., ``AdBlock'', ``Adblock for Youtube'', or ``AdGuard''), 1 enables web page layout customization (``Tampermonkey''), 1 performs grammar checks (``Grammarly''), 1 offers translation options (``Google Translate''), and 1 adds PDF editing capabilities (``Adobe Acrobat''). The 10\textsuperscript{th} extension is from the ``Shopping'' category (``honey''); it looks for and applies digital coupons while shopping online.

\begin{figure}[t]
    \centering
    \includegraphics[width=0.8\columnwidth]{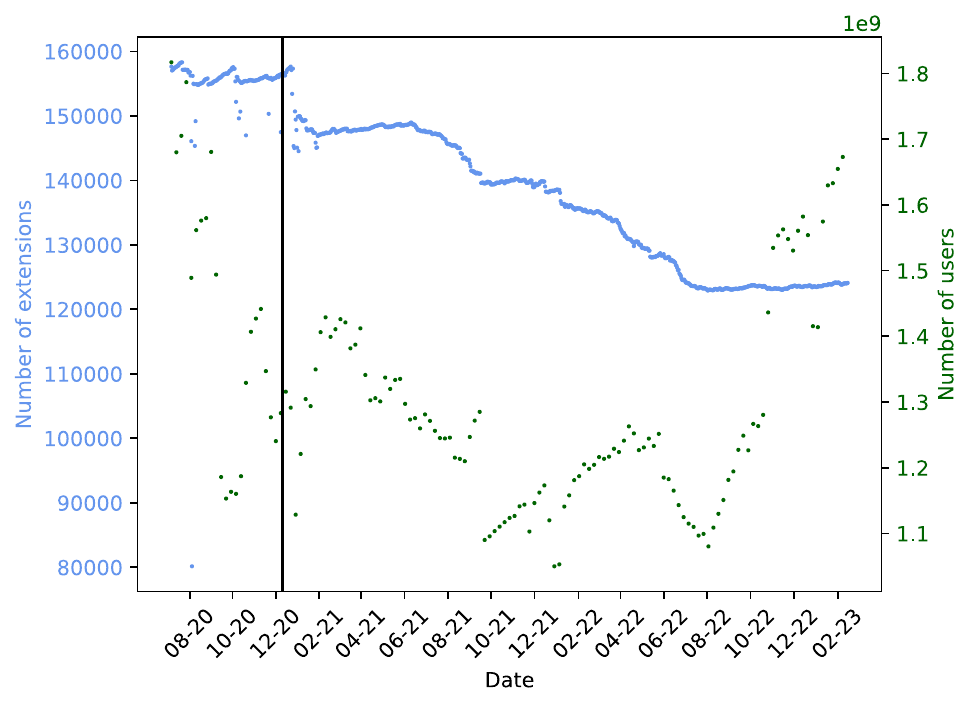} 
    \caption{Number of extensions (blue) in the CWS on a given date and weekly average of the number of active users across all extensions (green) between July 2020 and February 2023%
    \textnormal{--The black vertical line denotes the announcement of Manifest V3 on Chromium Blog on December 9, 2020}
    }
    \label{fig:num_exten}
\end{figure}

\pseudoparagraph{Number of Users}
We now examine the evolution of the number of extension users over time. The user count was scraped using the CWS API~\cite{Developers:vy}. This API provides exact user counts for all extensions, except for extensions with over 10 million users, which are just listed as having 10,000,000+ users.
According to our email exchange with Chrome Web Store Developer Support, the ``number of users'' displayed on the CWS for a given extension corresponds to ``the number of Chromes with the extension installed that are active and checking in to [their] update servers over the previous seven days only, not for all time. It is not equal to the sum of historic installs minus the sum of historic uninstalls''.
For instance, this means that if a user with an extension installed did not turn on their computer for a month, they would not be counted as an active user for that month.
Given this definition, we can therefore expect fluctuations in the number of active users over time.
We observe in \Cref{fig:num_exten} (green curve) that the number of extension users largely fluctuates over time. The exact reasons for this phenomenon are currently unclear, but we can make some observations. Overall, we see a large drop in the number of users during the holiday seasons (the last two weeks of December and the first week of January), which can be explained by people not using Chrome for over seven days as they spend time away from work. We also observe an overall dip in the number of users during the summer months, which could be caused by increased travel and idle school devices.

\summary{Takeaway}{There is unexpected volatility in the number of active users in the CWS. We caution researchers to not overly emphasize precise user counts when measuring the impact of work. We also recommend developers take this into account when reporting or comparing user counts over time.}

Regarding the average number of users per extension, we find that the vast majority of extensions have a small user base: 64.22\% of extensions have less than 100 users and 17.51\% have between 100--1k users. The higher the user count, the fewer extensions there are: 9.70\% of extensions have 1k--10k users, 4.17\% 10k--100k, 1.13\% 100k--1M, and 0.27\% over 1M.
Overall, extensions with a large user base are the exception rather than the rule.
This is problematic from a privacy perspective, as the crowd anonymity of extensions with few users is very small, which can ease user tracking and deanonymization if those extensions can be fingerprinted~\cite{Karami2020}.

\pseudoparagraph{Number of Extensions} \label{subsec:nb-be}
As shown in \Cref{fig:num_exten} (blue curve), the number of extensions is decreasing over time; notably with an initial 157k extensions in July 2020 vs.\ 124k in February 2023.
We observe a large drop in the number of extensions in December 2020--likely related to Chrome's November~9\textsuperscript{th} announcement of Manifest V3~\cite{manifestv3} (vertical black line).
Note that there were some (now resolved) bugs in the ChromeStats crawler when the developers released it in July 2020--as evidenced by some isolated points in the first 6 months of release and confirmed by the developers.

We observe that, on average, 3,775 extensions are removed from the CWS every month and another 2,687 are added. This information leads us to believe that, depending on the point in time of measurement, researchers may analyze different extensions; thus report on different results.

\summary{Takeaway}{Every month thousands of extensions are added or removed from the CWS. Given those fluctuations, we encourage researchers to investigate multiple points in time and evaluate the reproducibility of their findings before generalizing their claims.}

\subsection{Extension Life Cycle}

Very little is currently known about the lifetimes of extensions in the CWS.
It is important to investigate, as it sets guidelines for how long study results are valid. 
Analyses should be run regularly to ensure that research results reflect the current state of the CWS.

\begin{figure}[t]
    \centering
    \includegraphics[width=0.69\columnwidth]{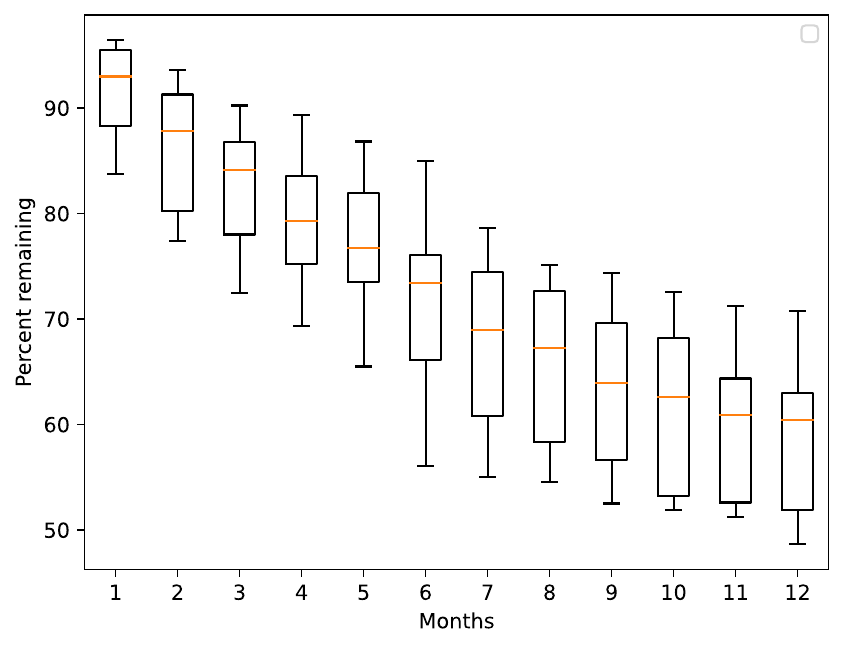}
    \caption{Percentage of extensions still in the CWS on the $x$\textsuperscript{th} month after having been added%
    \textnormal{--We compute this for extensions first added between January--December 2021 and still in the CWS on the $x$\textsuperscript{th} following month(s), where $x \in \llbracket 1, 12 \rrbracket$
    }}
    \label{fig:remain_month}
\end{figure}

We study extension life cycles by computing the percentage of extensions that remain in the CWS after a given number of months. To obtain a full year of data, we perform this analysis on extensions that were added to the CWS between January and December 2021, and we compute the percentage of those extensions that were still in the CWS after $x \in \llbracket 1, 12 \rrbracket$ months.
\Cref{fig:remain_month} represents the percentage of extensions still in the CWS on the $x$\textsuperscript{th} month after having been added, with $x \in \llbracket 1, 12 \rrbracket$.
For example, we consider extensions first added in January 2021 and still in the CWS in February 2021 ($x=1$), March ($x=2$), [...], and January 2022 ($x=12$). At the same time, we also study extensions first added in February 2021 and still in the CWS in March 2021 ($x=1$), April ($x=2$), [...], and February 2022 ($x=12$). We iterate this process till December 2021 (extensions added to the CWS) and calculate how many were in the CWS after 1--12 months (i.e., from January 2022 to December 2022).
This way, we have 12 data points for each 12 $x$ values (x-axis), which we represent with a box plot.
As shown in \Cref{fig:remain_month}, we find that after $x$ = 12 months, only 51.86--62.98\% of extensions are still available in the CWS (median of 60.39\%). Even after $x$ = 3 months, we observe a drop, with only 78--86.75\% of the extensions still available in the CWS (median of 84.11\%).

\begin{figure}[t]
    \centering
    \includegraphics[width=0.7\columnwidth]{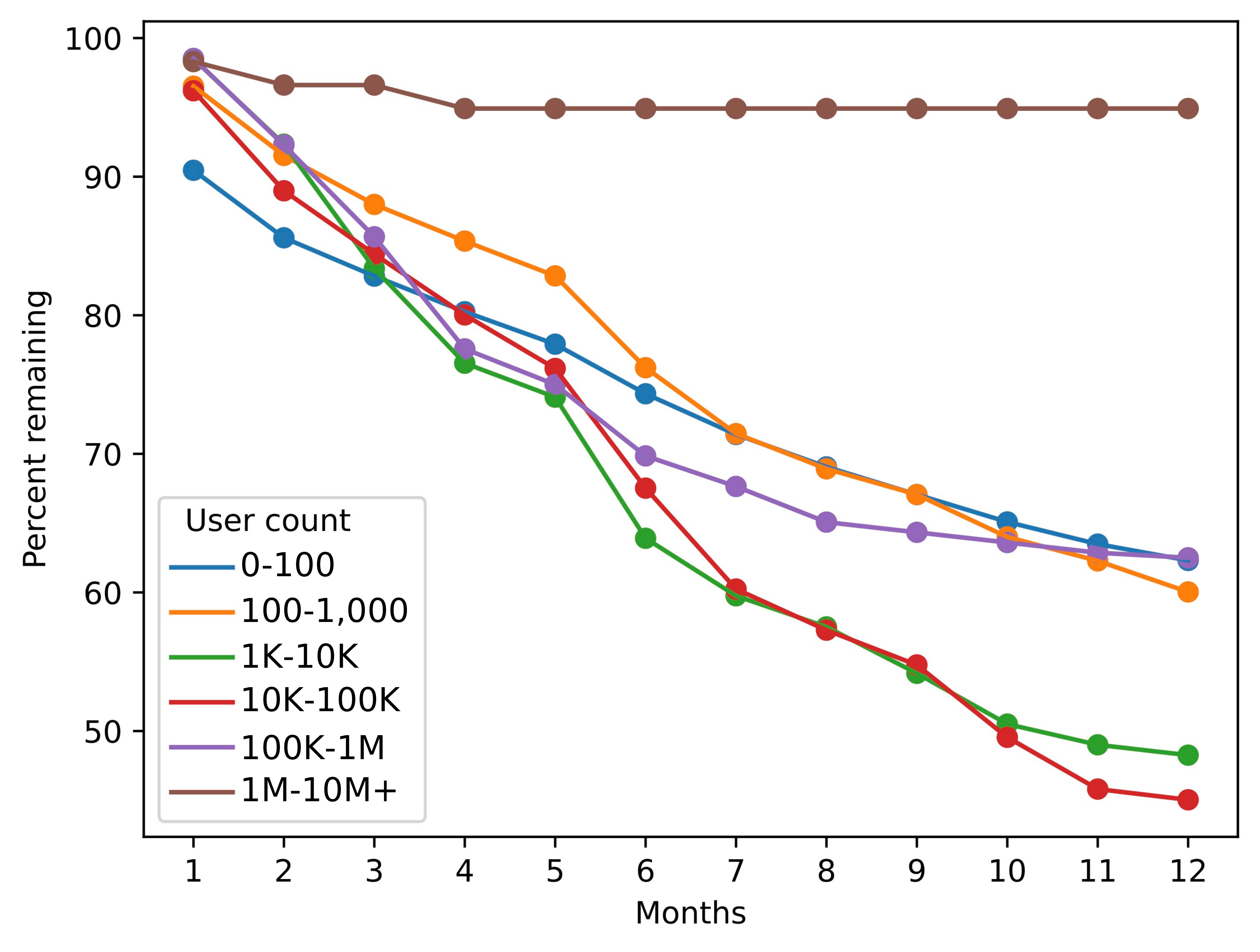}
    \caption{Average percentage of extensions still in the CWS on the $x$\textsuperscript{th} month after having been added, grouped by an extension last recorded user count}
    \label{fig:remain_buck}
\end{figure}

Intuitively, we expect extensions with a larger user base to remain in the CWS for a longer period, as it takes time to build up a large user base, i.e., developers are more likely to continue maintaining those extensions.
In \Cref{fig:remain_buck}, we show the average percentage of extensions still in the CWS on the $x$\textsuperscript{th} month after having been added, grouped by last recorded extension user count (i.e., so that extension groupings do not change over time).
We see that, on average, 94.91\% of the most popular extensions (1M--10M+ users) stay in the CWS for at least a year.
However, for the rest of the extension groupings, there is no clear trend. For example, the two groups with the highest percentage of extensions remaining after one year are extensions with less than 100 users (62.31\%) and 100k--1M users (62.5\%).
Upon further investigation, we find that there are numerous instances of popular extensions appearing in the CWS, being removed, then reappearing in the CWS. This pattern can be observed for extensions violating the CWS policies: they are removed by Google but can be submitted again after the violation has been fixed~\cite{ext-removed-added}.
In addition, there may be larger variability in the higher user count categories due to different sample sizes; e.g., 19,528 extensions have 100 users or less\footnote{Note that we conducted this analysis on extensions added between January and December 2021, not on our full extension dataset.} vs.\ 272 have 100k--1M users and 59 have 1M--10M+ users.

\summary{Takeaway}{Surprisingly, we observe that extensions have a very short life cycle in the CWS, e.g., only 51.86--62.98\% of extensions are still available after one year. We also observe instances of popular extensions that are added to the CWS, removed, and re-added; likely due to policy violations and subsequent fixes. Thus, analyses on the CWS should be run regularly to ensure that research results reflect the current state of the CWS.}

\section{Security-Noteworthy Extensions}

Previously, we focused on overall trends in the CWS, independently of whether the extensions were benign or security noteworthy.
We now dive into the differences between benign, malware-containing, policy-violating, and vulnerable extensions.
Specifically, we compare the number of days in the CWS, user counts, extension ratings, developer patterns, and permissions of these 4 groups of extensions.

\subsection{Number of Days in the CWS}

 \begin{figure}[t]
 		\centering \includegraphics[width=.8\columnwidth]{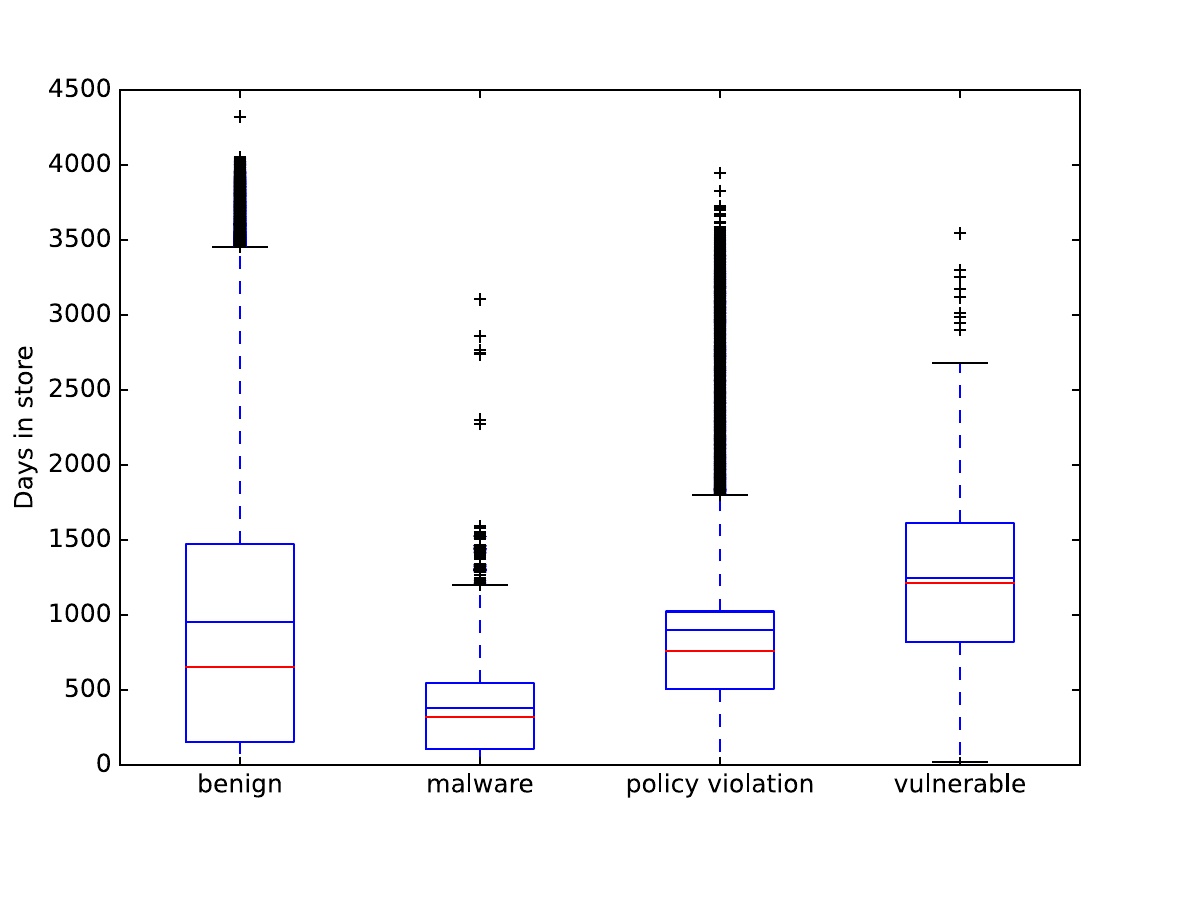}
 		\caption{Number of days a benign, malware-containing, policy-violating, or vulnerable extension stays in the CWS%
        \textnormal{--The blue line denotes the means and the red one the median}
        }
    \label{fig:days_in_store}
 \end{figure}

We first analyze the average number of days a benign extension vs.\ a SNE stays in the CWS.
For every malware-containing and policy-violating extension, we count the number of days between the date of its last update and the date of its removal. For vulnerable extensions, we count the number of days the reported vulnerable versions stayed in the CWS (as of May 1, 2023).
Of course, it is possible that an extension had security or privacy issues \textit{before} the last update or continues to be vulnerable even after being updated. We acknowledge that our observations are a \textit{lower bound} of the number of days SNE stay in the CWS.
As shown in \Cref{fig:days_in_store}, SNE stay in the CWS for an average of 380 days (malware containing) to 1,248 days (vulnerable). This is extremely problematic, as such extensions put the security and privacy of their users at risk \textit{for years}. Interestingly, benign extensions tend to stay in the CWS for less time than vulnerable extensions (1,152 days, with a median of 780 days vs.\ 1,213 for vulnerable extensions).
While the sample sizes are different (226,762 benign vs.\ 184 vulnerable extensions), there are more fluctuations within benign extensions.
Equally worrisome are some outliers. In particular, we found one malware-containing extension that stayed in the CWS for 3,105 days (8.5 years!). This extension, ``TeleApp'', was last updated on December 13, 2013 and was found to contain malware on June 14, 2022.
Similarly, the extension ``No More Holidays'' was last updated on May 17, 2012 and was found to have a policy violation only on March 9, 2023 (after almost 11 years in the CWS!).

\summary{Takeaway}{Security-noteworthy extensions stay in the CWS \textit{for years}, meaning that their user base can stay at risk \textit{for years}. It is currently unclear why such extensions are not immediately detected by Chrome's vetting system and why SNE stay in the CWS for years even after disclosure (e.g., case of vulnerable extensions).}

\subsection{Number of Users}

\begin{figure}[t]
	\centering \includegraphics[width=0.8\columnwidth]{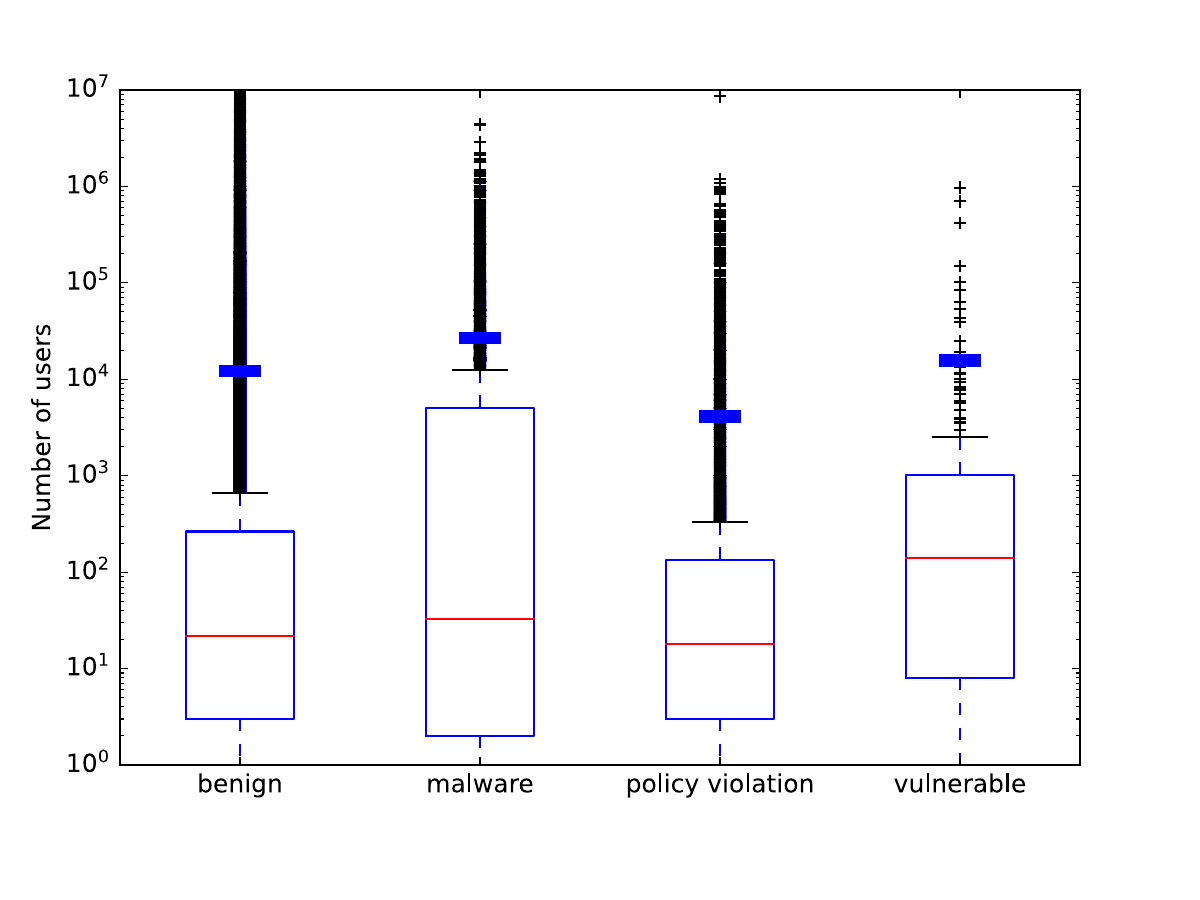}
	\caption{Number of users with a benign, malware-containing, policy-violating, or vulnerable extension installed%
    \textnormal{--The blue tick denotes the means and the red line the median}
    }
	\label{fig:all_four}
\end{figure}

Next, we investigate how large the user bases of SNE are. To this end, we consider the number of users they had when removed from the CWS or, if not yet removed, May 1, 2023 (when we conducted this analysis).
As a comparison, we also consider the last recorded number of users for benign extensions.
We represent the number of users of benign extensions and SNE in \Cref{fig:all_four}.
As expected, the median number of users is very low: between 18 (policy-violating and benign extensions) and 140 (vulnerable extensions); remember that 65\% of extensions have less than 100 users (\Cref{subsec:trends}).
However, there are some outliers with extremely large user bases so that, on average, benign extensions have 11k users, policy-violating 4k, vulnerable 16k, and malware-containing extensions 27k.
For example, the extension ``Casino Volcano'' had 8.58M users and was ranked number five in the ``Social and Communication'' category until it was removed on July 20, 2020 for policy violations. 
Quite worrisome is the fact that 25\% of malware-containing extensions have over 5k users (vs.\ a 75\textsuperscript{th} percentile of 220 for benign extensions).

In total, we collected the metadata of over 26k SNE (\Cref{tab:dataset-sne}). Overall, we observe that over 346 million users installed at least one SNE in the last three years: 280M users installed malware-containing extensions, 63.3M policy-violating, and 2.9M vulnerable extensions.\footnote{Note that we know the number of users for each extension, but we cannot deduplicate the total number of users, in case one user has several SNE installed; so this number does not represent \textit{unique} users.}
We assume that those users are unaware of using SNE. Given both the extremely large number of impacted users and the fact that SNE stay in the CWS \textit{for years}, SNE are a major problem and need to be removed as quickly as possible from the CWS. While Google engineers seem to be looking for malware-containing or policy-violating extensions through their review process~\cite{chrome-vetting} (with more or less success); to the best of our knowledge, they currently cannot detect vulnerable extensions~\cite{fass2021doublex, Yu2023}. Future work would be beneficial in this area to further secure the CWS.

\summary{Takeaway}{\textit{Over 346 million users} installed a SNE in the last 3 years. Given that SNE tend to stay in the CWS \textit{for years}, it is critical to improve the detection of SNE and notify impacted users.}

\subsection{Extension Ratings}

Given that SNE are only moderately removed from the CWS and that this process can take up \textit{years}, we now investigate if users themselves are able to flag SNE, e.g., by giving such extensions a low rating.
On the CWS, extension users can rate extensions from 1 (lowest score) to 5 (highest score); extensions with no ratings are given a score of 0.
First, we observe that a large number of extensions have no ratings: 58.63\% of policy-violating extensions have no ratings, 52.12\% for malware, 47.32\% for vulnerable, and 31.52\% for benign extensions.
Of the extensions that do have a rating, we do not observe a significant difference in scores between these four groups: the benign and policy-violation sets have a median of 5, 4.997 for malware, and 4.571 for vulnerable.
Overall, users do not give SNE lower ratings, suggesting that users may not be aware that such extensions are dangerous. Of course, it is also possible that bots are giving fake reviews and high ratings to those extensions.
However, considering that half of SNE have no reviews, it seems that the use of fake reviews is not widespread in this case.

\summary{Takeaway}{Users do not give SNE lower ratings. This makes thorough vetting and review of extensions all the more critical.}

\subsection{Extension Developers} \label{subsec:dev}

\begin{figure}[t]
	\centering \includegraphics[width=0.635\columnwidth]{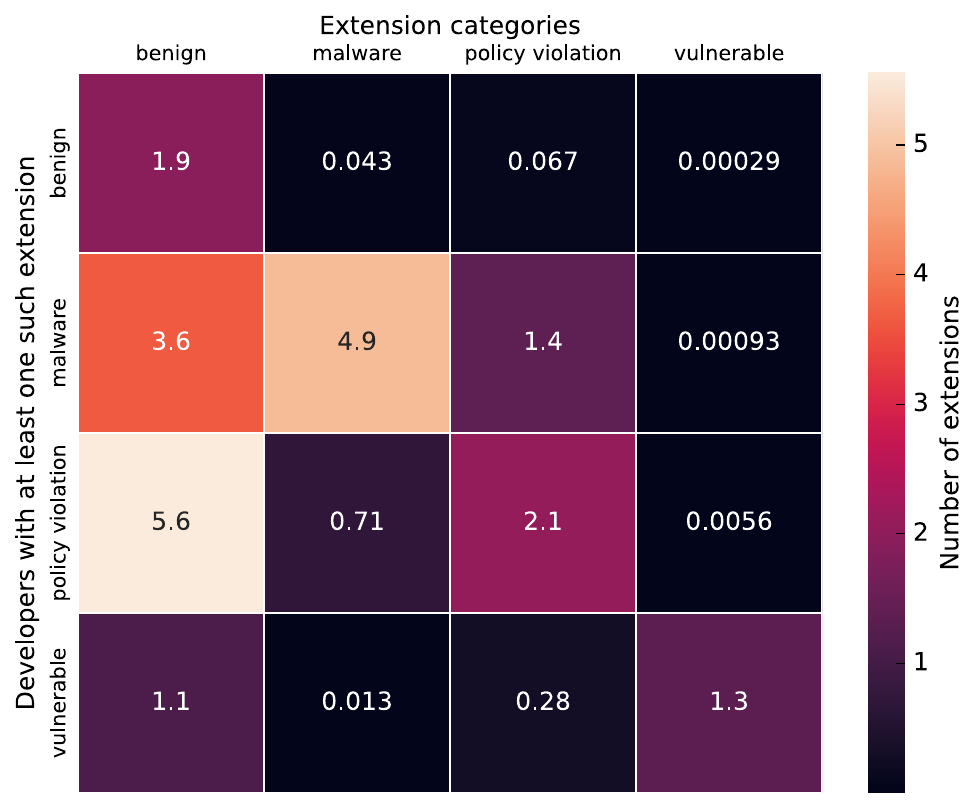}
	\caption{Average number of benign extensions or SNE published by a developer with at least one of these extensions%
 \textnormal{--E.g., a developer with 1 malware-containing extension in the CWS (y-axis) publishes on average 3.6 benign, 4.9 malware-containing, 1.4 policy-violating, and 0.00093 vulnerable extensions (x-axis)}
 }
	\label{fig:developers}
\end{figure}

Next, we focus on the developers designing the SNE that are or were in the CWS.
Interestingly, we found numerous cases of developers publishing both SNE and benign extensions.
In \Cref{fig:developers}, we represent the number of benign, malware-containing, policy-violating, and vulnerable extensions (x-axis) published by a developer already having one such extension in the CWS (y-axis).
For example, a developer having published 1 malicious extension (y-axis) publishes on average 3.6 benign, 4.9 malware-containing, 1.4 policy-violating, and 0.00093 vulnerable extensions.
We also found instances of developers having many malicious extensions. For example, the developer ``http://newtabexperience.com'' has had 1,041 extensions removed for containing malware and 9 for policy violations (as of May 1, 2023, this developer still has 434 extensions in the CWS).
Overall, we observe that developers with at least one SNE tend to publish more SNE than developers with one benign extension. In particular, developers publishing malware-containing extensions publish almost 5 of those, on average, whereas other developers publish less than 1. In other words, it is quite unlikely that a developer having published at least one benign extension will publish a SNE.
There are some exceptions, though. For example, the developer ``New Tab'' has 1,041 benign extensions
in the CWS but also 2 SNE (1 malicious and 1 policy-violating extension).
On the contrary, a developer having a malware-containing or privacy-violating extension will likely publish another one of those.
Overall, there are 30 developers with over 100 malware-containing extensions each; and 28 developers each of them with over 100 extensions removed for violating the CWS policies. 
We assume that such developers multiply their SNE to try to infect or harm as many users as possible, for their own profit. Therefore, extensions by developers known to have created a SNE may require further scrutiny.
The trend is different for vulnerable extensions, though. Such developers do not seem to publish more malware-containing or privacy-violating extensions than benign developers, but they tend to publish more vulnerable extensions. So, contrary to the other SNE, it seems that developers publishing vulnerable extensions are \textit{well intentioned} but make mistakes in their implementation, leading to \textit{vulnerabilities}.

\summary{Takeaway}{Developers having published a SNE are more likely to publish SNE than developers having published a benign extension. We endorse flagging and further scrutinizing the extensions submitted to the CWS by developers with known SNE.}

\subsection{Extension Permissions}

\begin{figure}[t]
    \centering
    \includegraphics[width=0.7\columnwidth]{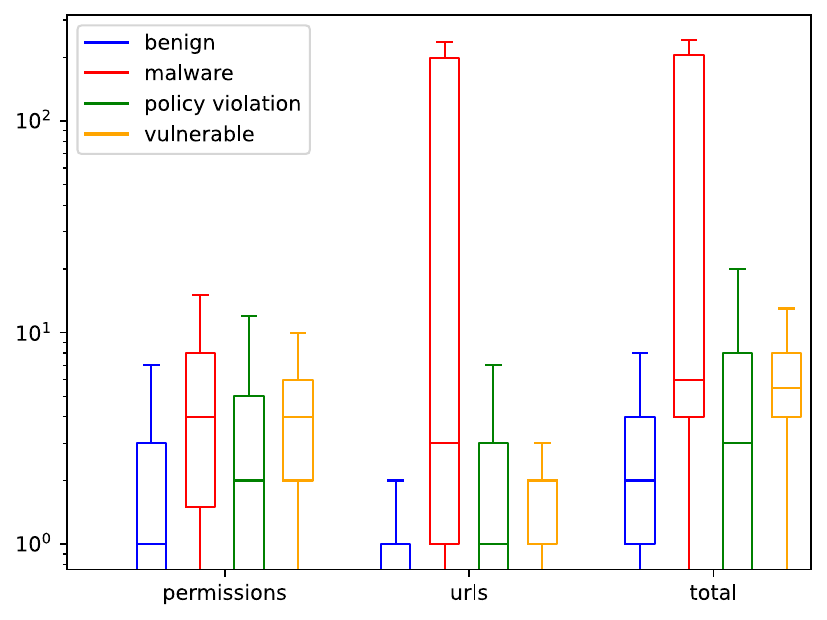}
    \caption{Number of API-based (permissions), host-based (urls), and total (sum of APIs and hosts) permissions of benign, malware, privacy-violating, and vulnerable extensions}
    \label{fig:num_permissions}
\end{figure}

Finally, we compare the permissions a SNE vs.\ benign extension request. In fact, the requested permissions will determine the capabilities of an extension, i.e., the attack surface for SNE.
We collected permissions by parsing each extension's \texttt{manifest.json} file.
For manifest V3, permissions are broken up into \textit{permissions} (APIs such as \texttt{storage} or \texttt{cookies}) and \textit{host permissions} (URLs or URL patterns that an extension wants to make requests to).
For manifest V2, these two categories are combined. For consistency with manifest V3, we separate APIs from host permissions.

First, we look at the number of permissions each extension requires (\Cref{fig:num_permissions}).
As expected, SNE require more permissions than benign extensions.
For API-based permissions, malware-containing and vulnerable extensions use a median of 4 permissions vs.\ 2 for policy-violating and 1 for benign extensions.
We observe a similar trend for host permissions. If we look at the third quartile, 25\% of malware-containing extensions list 198 URLs; 3 for policy-violating, 2 for vulnerable, and 1 for benign extensions.
Ultimately, the more permissions an extension has, the larger the attack surface is.

\begin{table}[t]
\centering
\resizebox{\columnwidth}{!}{%
\begin{tabular}{l | l l l l l l}
\toprule

Rank   &  Benign & Malware & Policy violation & Vulnerable  \\
\midrule
1 & <all\_urls> & *://www.google.com.kh/* & *://www.google.com.kh/* & http://*/*\\
2 & https://*/* & *://mail.google.com/* & *://mail.google.com/* & https://*/*\\
3 & http://*/* & *://www.google.com/* & *://www.google.com/* & <all\_urls>\\
4 & *://*/* & *://www.google.com.au/* & *://www.google.ps/* & chrome://favicon/\\
5 & http://*/ & *://www.google.us/* & *://www.google.co.zw/* & *://*.fliptab.io/*\\
6 & https://*/ & *://www.google.nl/* & *://www.google.co.zm/* & *://*/*\\
7 & chrome://favicon/ & *://www.google.ca/* & *://www.google.co.za/* & https://*/\\
8 & https://ajax.googleapis.com/ & *://www.google.dk/* & *://www.google.ws/* & http://*/\\
9 & https://image.lovelytab.com & *://www.google.co.jp/* & *://www.google.vu/* & *://127.0.0.1/*\\
10 & https://v2.lovelytab.com/ & *://www.google.no/* & *://www.google.com.vn/* & *://localhost/*\\

\bottomrule 
\end{tabular}
}
\quad

\caption{Top 10 host-based permissions for benign and SNE}
\label{table:urls_top}
\end{table}

Next, we dive into the specific API and host permissions extensions need.
Interestingly, and perhaps surprisingly, both benign extensions and SNE seem to more or less use the same APIs (see a list of the top 10 API-based permissions in \Cref{table:permissions_top} in the Appendix). The only notable exception is \texttt{topSites} (permission allowing an extension to access a user's most visited sites), ranking 2 for malicious extensions (4,026 extensions) but not in the top 10 in any of the other groups.
This number is likely influenced by the numerous malware-containing extensions that replace a user's homepage on new tabs, which requires this \texttt{topSites} permission to operate.

On the contrary, there are differences between the host permissions required by each category of extensions, as illustrated in \Cref{table:urls_top}.
Firstly, all-encompassing URLs, e.g., ``<all\_urls>'' or ``http://*/*'', are highly ranked in both benign and vulnerable extensions but not in malware-containing nor policy-violating, which instead have a lot of Google sub-domains.
One reason for this may be that extensions requesting powerful privileges, such as ``<all\_urls>'', require in-depth review~\cite{chrome-vetting}, which developers with malicious intent may want to avoid.
For example, we discovered a set of 1,130 SNE, 740 of which were removed for malware and 390 for policy violation; all of them had a large list of Google sub-domains in their host permissions.
This exemplifies how large clusters of identical extensions can skew results and demonstrates the need to account for duplicated extensions when conducting analyses.

\summary{Takeaway}{SNE request more permissions than benign extensions, likely because the more permissions an extension has, the larger the attack surface is. While host-based permissions differ between benign and malware-containing extensions, API permissions do not enable to discriminate between benign and SNE.}

\section{Code Similarity and Security}

Next, we show that analyzing extension codes for similarity could be a useful tool for identifying SNE.
Subsequently, we investigate the sources of code similarities and exemplify how code reuse can lead to the propagation of vulnerabilities.

\subsection{Extensions with a Similar Code Base} \label{subsec:similar-code}

We first present our approach to detect clusters of similar extensions and discuss some case studies.

\subsubsection{Methodology} \label{subsubsec:similar-code-methodo}
As described in \Cref{subsec:ext-collection}, we downloaded and unpacked our sets of malware-containing, policy-violating, vulnerable, and benign extensions.
For each extension, we extracted their content and background scripts, as described previously.
We could successfully collect 108,859 extensions (92,482 benign, 6,587 malware-containing, 9,638 policy-violating, and 152 vulnerable).
To compare the source code of all these extensions with each other, we compared all pairs of \texttt{(content script, content script)}, \texttt{(background script, background script)}, and \texttt{(background script, content script)}.\footnote{We found clusters of similar (content script, background script). E.g., the background script of ``PodQueue'' is similar to the content script of ``Slack Beeegmojis'', as they both use the \texttt{webextension-polyfill} class in their respective scripts.}
For all of the described pair types, we performed an ssdeep fuzzy-hash-based clustering~\cite{ssdeep}.
We chose this similarity-hashing algorithm to identify clusters of similar%
\footnote{We prefer the word \textit{similar} to \textit{identical}: while the majority of the scripts are identical word for word, a few scripts have the same syntactic structure, but a few string constants were changed. We hypothesize they were still clustered with a 100\% overlap, because ssdeep returns an integer between 0 (no match) and 100, and rounding up by the algorithm probably gave a 100 score to \textit{nearly} identical extensions.}
extension scripts.
We cluster two extensions together if they have a 100\% overlap either with their content script or background script.

\subsubsection{Clusters of Similar Extensions}
We compared the source code of all 108,859 extensions we successfully collected. We found that 20,822 extensions contain a similar content or background script to other extensions.
We found a total of 3,270 clusters. The median cluster size is 2 similar extensions, but the distribution is heavily right-skewed with a mean of 7.32 extensions and a maximum cluster size of 1,397 extensions.
By manually analyzing extension source code and CWS pages, we categorize our clusters into three types:

\begin{itemize}[leftmargin=*]

     \item {\bf Identical extensions:}
     These extensions look the same and perform the same tasks; their description and images in the CWS are similar.
     E.g., we observe instances of developers publishing a new extension instead of pushing an update to an existing one.
    
     {\bf Case study:} We found 19 ``Kobo Book'' extensions (e.g., ``Free Kobo Books'', ``Free Kobo Books Deluxe'', or ``Free Kindle Religious Books''). They all have identical background scripts as well as identical screenshots and descriptions on their respective CWS pages, but some have different names and developers.

 \vspace{+0.1cm}
     \item {\bf Extensions with a similar functionality:}
     These extensions have a similar functionality, but they may look different and have a dissimilar page on the CWS. E.g., there are many home page wallpaper extensions, where a developer publishes hundreds of extensions with a different theme, such as cars or nature. The source code for all of these extensions is the same, with a different URL string in the code to select the specific image category.

     {\bf Case study:} 9 screenshotting extensions share an identical background script, e.g., ``Xsnap'', ``LTR Screenshot'', or ``Computub Screenshotter''. Although the extensions do similar actions, they have very different UI and branding. The identical background scripts contain code to receive a button click, take a screenshot, and display it to the user.

 \vspace{+0.1cm}
     \item {\bf Extensions with generic background scripts:}
     Such extensions perform very different functions but contain generic code, such as receiving a message or sending a response, which could come as part of an extension builder or sample code.
     
      {\bf Case study:} We found a cluster of 1,397 extensions with identical background scripts, e.g., ``Read a Newspaper'' or ``Goleferine''. The background script for all of these extensions is a message listener that sends the tab ID. These files are identical, down to the comment ``//example of using a message handler from the inject scripts,'' which makes it likely that all developers copied the code from a Chrome tutorial or StackOverflow post.

 \end{itemize}

\subsubsection{Suspicious Extension Clusters} \label{subsubsec: susp-clusters}

Of the 3,270 clusters we identified previously, 2,296 contain only benign extensions, 321 only SNE, and 653 both SNE and benign extensions.
The fact that 321 clusters contain only SNE is a good indication that code similarity could enable to quickly remove a full cluster
once one of the extensions in the cluster is found to be security critical; or at least to flag the remaining extensions in the cluster for further analyses.
In the following, we focus in more detail, first on SNE clusters, and then on clusters containing both SNE and benign extensions.

\pseudoparagraph{SNE Clusters}
Regarding SNE clusters, 14 of those contain 100 or more SNE, with 2 clusters containing 863 SNE each.
One of these 863-SNE clusters mainly contains new tab wallpaper extensions, i.e., when a user opens a new tab, such extensions redirect a user to a custom Google search page with a different wallpaper image.
These wallpaper images can come in different themes, e.g., sports or anime, which allows a developer to create hundreds of extensions with the same code but different background images.
Such extensions include ``Kpop Big Bang Wallpapers New Tab HD'' and ``Los Angeles Clippers Wallpapers New Tab HD''.
The majority of these new tab extensions are by the developer ``http://newtabexperience.com'', although there are extensions by other developers in the cluster. Across the board, new tab extensions are very common, as they can be used to show users advertisements or distribute malware~\cite{Majauskas:wc}.
In fact, almost all of the extensions in this cluster were removed from the CWS for containing malware on October 1--5, 2020. 

\pseudoparagraph{SNE and Benign Extension Clusters}
We now look at the 653 clusters containing both benign extensions and SNE. These clusters total 10,678 extensions: 1,754 malware-containing, 3,370 policy-violating, 2 vulnerable, and 5,552 benign extensions.

One such cluster contains 9 screenshotting extensions that all have the same background script.  3\,/\,9 extensions were removed for policy violations months apart (January 20, February 11, and March 17, 2023), despite them having \textit{identical code} and not receiving any updates in this time span.
As a result, we believe that the CWS developer team is not using code similarity as part of their review process to help detect suspicious extensions. With code similarity-based approaches, once one extension in the cluster had been found to be in violation of the CWS policies, the other 8 identical extensions could have been flagged for additional review, allowing a faster removal of those policy-violating extensions.

Interestingly, we also found examples of truly benign extensions sharing a cluster with SNE.
For example, in a cluster of 9 extensions (7 benign and 2 malicious), we see that ``Keplr for cosmos: osmosis,akt Wallet'' is malicious but not ``G-Eye''. Despite them having \textit{identical code}, they have \textit{different permissions}: the benign extension is missing the \texttt{activeTab} and \texttt{scripting} permissions. This makes the ``malicious'' part in the benign extension dead-code for now (due to the absence of the required permissions).
In other words, a simple permission change in the \texttt{manifest.json} file could make the extension malicious.
Unfortunately, both Fass et al.\@~\cite{fass2021doublex} and Pantelaios et al.\@~\cite{Pantelaios2020} reported examples of extensions turning vulnerable or malicious, respectively, after such an update.
For these reasons, we believe that the 5,552 ``benign'' extensions containing \textit{identical code to known SNE} should be flagged for additional review. Maybe some of them are not security critical today, but they could easily become so, and should thus be more closely monitored.

\summary{Takeaway}{We uncovered thousands of clusters of similar extensions. We found that 321 clusters contain only SNE, showing that analyzing extension code for similarities could be a useful tool for identifying SNE: once one extension in a cluster is found to be security-critical, researchers or Google developers could flag the remaining ones for additional screening.
Flagged extensions could then be run through more computationally and time-expensive analyses, and should be more closely monitored.}

\subsection{Sources of Code Similarity}

Interestingly, some of the similar extensions we found are written by different developers.
We observe that it is common for developers to include URLs to reference the origin of a code snippet or give credit to other authors~\cite{hata20199}.
As a best-effort strategy to understand sources of code similarity, we scrape the comments from all of the downloaded extension code (JavaScript files) and extract URL-like strings. While sometimes the URLs are simply generic license files, we find that in many cases links cited in commented code specify the origin of the code, i.e., the reason for code similarity.

\subsubsection{Methodology} 
For each extension, we extract the comments from all of the JavaScript files using Esprima \cite{esprima}, a popular JavaScript parser used by prior work~\cite{Staicu2018SYNODEUA, Salih2016, fass2019hns, fass2019jstap, fass2018jast, fass2021doublex, Soni2015, Han2019, Lee2020, moog2021}.
To conduct a more fine-grained analysis, contrary to \Cref{subsec:similar-code}, we consider all JavaScript files from an extension package independently (i.e., we do not merge all content scripts or background scripts).
Then, we use the get-urls library to search for and extract URLs from the comments~\cite{getUrls}. Since developers also use comments to temporarily remove small sections of code, the get-urls library triggers false positives (i.e., text being incorrectly flagged as URL).
To eliminate these false positives for our analysis, we filter for strings that contain one of the top five most popular top-level domains: ``.com'', ``.org'', ``.ru'', ``.net'', and ``.uk''~\cite{popTLD}. We also include ``.io'' and ``.edu''.

\subsubsection{Origins of Code Similarity} 
Overall, we collected over 8M URLs embedded in comments, over 50k of which are unique (we list the top 10 URLs appearing the most often in \Cref{table:esprima_url} in the Appendix).
These URLs belong to 18,551 unique second-level domains (we list the 10 most popular second-level domains in \Cref{table:esprima_rootdomain_unique} in the Appendix). As a best-effort strategy to understand to which categories of websites these URLs point to, we manually inspected the top 100 most popular URLs.
We find URLs primarily from the following categories: licenses, code documentation, code repositories, forum responses, corporate websites, and personal developer websites.
Overall, we collected 697 unique URLs that contain licenses, 3,036 from Stack Overflow, and 9,014 from GitHub repositories.

Next, we validate our assumption that identical URLs in code commented blocks are an indicator of code similarity.
To this end, we leveraged the ssdeep library~\cite{ssdeep}, as previously, to compute the fuzzy hashes of all the files that contain a given URL in their commented code, and we compared all the hashes.
To be able to manually verify our findings, we used this approach on a random sample of 100 extensions, for each of the ten most popular URLs, as listed in \Cref{table:esprima_url_similarity}, columns 1 and 2.
In column 3, we report the proportion of extensions that have a unique file. For example, 45\% of the extensions listing the URL ``\url{http://apache.org/licenses/LICENSE-2.0}'' do not have any files identical to any files from the other 99 extensions listing this URL. This result is not particularly surprising as this URL refers to a license file, which can be applied to a variety of different software and is, therefore, not a strong indication of code similarity.
On the contrary, for ``\url{http://extensionizr.com}'', only 15\% of the files referencing this URL have unique files not appearing in any of the other 99 extensions.
Overall, the results from \Cref{table:esprima_url_similarity} confirm the fact that extensions reuse code from public sources, such as code documentation, code repositories, or forums.
This is bad practice, as code reuse multiplies the usage of dated code and propagates vulnerabilities~\cite{Pearce2022}. We give a specific example below.

\subsubsection{Security Implications}
In this section, we illustrate the security implications of code reuse with one case study. We focus on Extensionizr, an open-source project used for generating boilerplate for Chrome extensions~\cite{extensionizr}.
As shown in \Cref{table:esprima_url_similarity}, ``\url{http://extensionizr.com}'' is the fourth most frequent URL found in at least one comment block of an extension.
Despite Extensionizr being last updated in 2017~\cite{extensionizrGithub}, 816 new extensions have been created using this project between 2020 and 2023 (we plot the number of extensions created each year between 2012--2023 using Extensionizr in \Cref{fig:extensionizr} in the Appendix).
Additionally, extensions created with Extensionizr have the option of adding Angular.js or jQuery as add-ons. However, since the project is not maintained, the only options for adding these libraries are Angular.js version 1.0.6 and jQuery version 2.0.0, both of which have known vulnerabilities~\cite{retireJSVulnerable}.
To estimate a lower bound of the number of Extensionizr extensions using a vulnerable library, we used RetireJS, an open-source tool to detect known vulnerable libraries~\cite{retireJS}.
We found that 65.56\% of the Extensionizr generated extensions use the default and vulnerable jQuery 2.0.0 and 79.66\% Angular.js 1.0.6 (note that those vulnerable libraries are not necessarily exploitable from the context of a web application).
This case study rather exemplifies the fact that code reuse is a bad practice that can lead to dated code and vulnerabilities, all the more when the third-party code is not maintained.

\summary{Takeaway}{We find that commented URLs in extension code can indicate the origins of code similarity. Code reuse is a bad practice, as it propagates the usage of dated code and vulnerabilities. For instance, roughly 1,000 extensions use the open-source Extensionizr project, 65--80\% of which still use the default and vulnerable library versions initially packaged with the tool, 6 years ago.}

\section{Maintenance and Security}

Previously, we showed that dated code propagates vulnerabilities.
We now investigate extension maintenance in the CWS and discuss security implications. Specifically, we exemplify the prevalence of outdated and vulnerable developer practices.

\subsection{Extension Updates} \label{subsec: updates}

\begin{figure}[t]
    \centering
    \includegraphics[width=0.65\columnwidth]{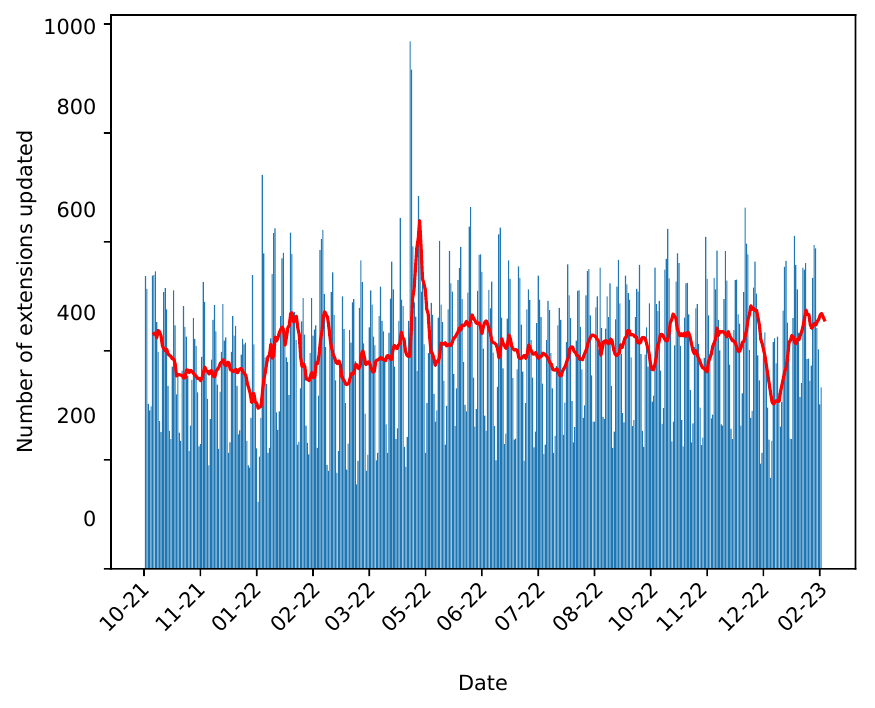}
    \caption{Number of extensions updated on a given day%
    \textnormal{--The red line represents the 7-day rolling average}}
    \label{fig:update_day}
\end{figure}

First, we focus on extension updates.
On average, 200--600 extensions are updated every day (\Cref{fig:update_day}). 
The number of updates per day was mostly constant over the past two years.
Interestingly, there were no update spikes after any key releases about the manifest V3 timeline. However, there are two outliers on April 19-20, 2022 where over 900 extensions were updated on each day. This correlates with the release of the featured and verified badges, which was reported by news outlets on April 19 and officially posted on Google's blog on April 20 \cite{Kim:ta}.
We assume that many developers submitted updates to meet the requirements for the badges.

\begin{table}[t]
\centering
\footnotesize
\resizebox{\columnwidth}{!}{
\begin{tabular}{l r r}
\toprule

URL & \begin{tabular}[c]{@{}l@{}}\# unique \\ extensions\end{tabular} & \begin{tabular}[c]{@{}l@{}}\% extensions\\ with a unique file\end{tabular} \\

\midrule
           http://apache.org/licenses/LICENSE-2.0 & 2,622 & 45\% \\
               http://opensource.org/licenses/MIT & 2,122 & 18\% \\
   http://opensource.org/licenses/mit-license.php & 1,735 & 74\% \\
                          http://extensionizr.com & 1,651 & 15\% \\
                        http://jquery.org/license & 1,158 & 21\% \\
                          http://underscorejs.org &  854 & 12\% \\
               http://code.google.com/p/crypto-js &  755 & 16\%  \\
  http://code.google.com/p/crypto-js/wiki/License &  754 & 20\% \\
        https://github.com/carhartl/jquery-cookie &  722 & 15\%  \\
                       http://mozilla.org/MPL/2.0 &  712 & 35\% \\

\bottomrule 
\end{tabular}
}
\quad 

\caption{Number of unique extensions a URL appears in and percent of extensions with this URL that have unique files
}
\label{table:esprima_url_similarity}
\end{table}

Surprisingly, the median number of updates an extension receives after a year in the CWS is 0, and the average is 0.16. So, the majority of extensions do not seem particularly maintained (even though some outliers have received over 25 updates).
Of the extensions in the CWS on February 14, 2023, 73,865 (59.5\%) have never been updated since they were added to the CWS.
Interestingly, those extensions tend to have fewer users (average of 1,974) than extensions that have received updates (30,708 users).

\begin{figure}[t]
    \centering
    \includegraphics[width=0.69\columnwidth]{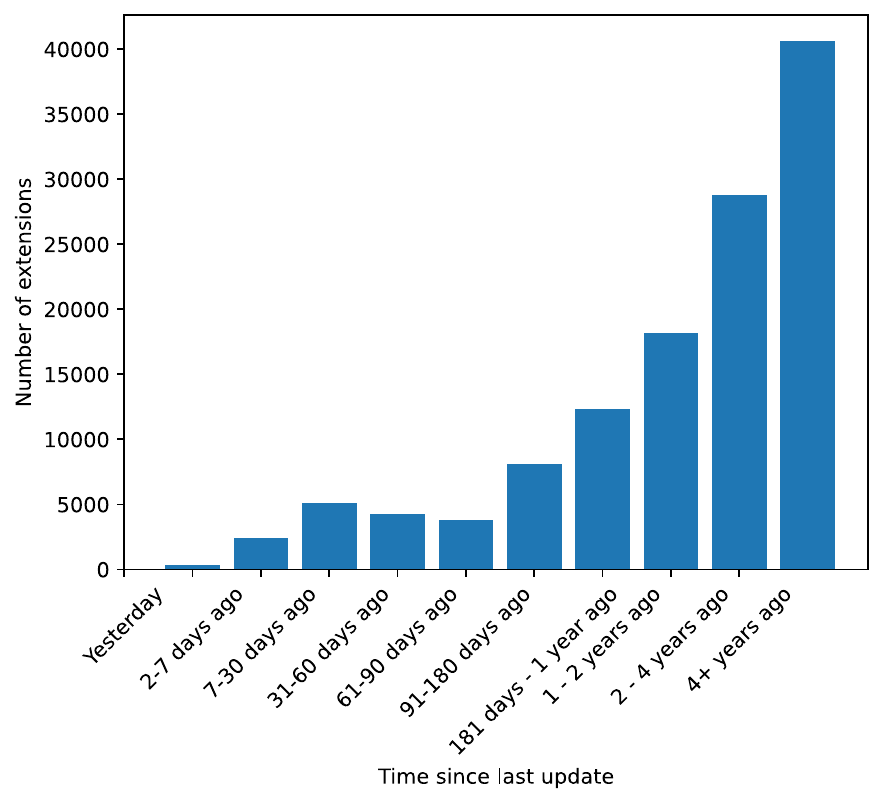}
    \caption{Number of extensions at their last update%
    \textnormal{ (as of February 14, 2023)}
    }
    \label{fig:updates_months}
\end{figure}

To further quantify our claim about the lack of maintenance in the CWS, we represent the number of extensions by time window of their most recent update in \Cref{fig:updates_months} (as of February 14, 2023). As expected, the majority of extensions have not been updated within the past year: almost 30k extensions have been last updated 2--4 years ago, and over 40k more than 4 years ago. We even observe over 5k extensions not having been updated within the past decade!
Overall, this raises concerns about compatibility issues and potential security or privacy implications of unmaintained extensions. We give a specific example in \Cref{subsec:lifecycle-vuln}, where half of known vulnerable extensions have been unmaintained for over 2 years.
Nevertheless, we acknowledge that some extensions may already have achieved their full functionality when being first added to the CWS and may not need further updates.
Still, Chrome is releasing new APIs and features, such as Manifest V3, whose benefits (also including security and privacy improvements~\cite{manifestv3-info}) those unmaintained extensions are missing out on.

\summary{Takeaway}{We highlight a critical lack of maintenance in the CWS, e.g., almost 60\% of extensions have never received any updates. Given the sensitive nature of extensions, this is problematic, as such unmaintained extensions are missing out on security and privacy improvements such as those offered by Manifest V3.}

\subsection{Manifest Versions}

As discussed in \Cref{sec:background}, Chrome released manifest V3 in November 2020 and stopped accepting new manifest V2 extensions in January 2022~\cite{manifest-v2}.
However, as of February 2023, 74\% of extensions are still using manifest V2 (62\% as of July 2023). This is especially concerning as Chrome originally planned to force all manifest V2 extensions to set their visibility to unlisted or private in June 2023 and remove all manifest V2 extensions in January 2024, although this is now under review and being postponed~\cite{manifest-v2}.
The purpose of manifest V3 is to improve security, privacy, and performance~\cite{manifestv3-info}.
For example, from a security perspective, extensions with manifest V3 are no longer able to run arbitrary code in their highly-privileged service worker environment. All code must be packaged within the extension, unlike previously where extensions could download and run external code~\cite{code-exec}. 
In particular, it has been shown that extensions running arbitrary code could lead to universal cross-site scripting vulnerabilities~\cite{Some2019, fass2021doublex, Yu2023}. These vulnerabilities (among others) would be fixed if the extensions migrated to manifest V3.

While manifest V3 promises a higher security and privacy posture, transitioning from V2 to V3 is challenging. For example, the transition requires significant changes for many existing extensions: some features are now unavailable, which forces developers to invest considerable time and effort to look for alternatives, which sometimes do not exist. In other cases, developers complain about limited API capabilities: the shift from the \texttt{webRequest} API to the \texttt{declarativeNetRequest} API limits extensions' ability to modify web requests on the fly, which is problematic, e.g., for adblockers~\cite{chromestats-v3}.

\summary{Takeaway}{Unfortunately, the transition from manifest V2 to V3 is challenging. As a consequence, 74\% of extensions and 1.2 billion users are potentially missing out on some security and privacy benefits offered by manifest V3.}

\subsection{Vulnerable Extensions}
\label{subsec:lifecycle-vuln}

\begin{figure}[t]
    \centering
    \includegraphics[width=0.669\columnwidth]{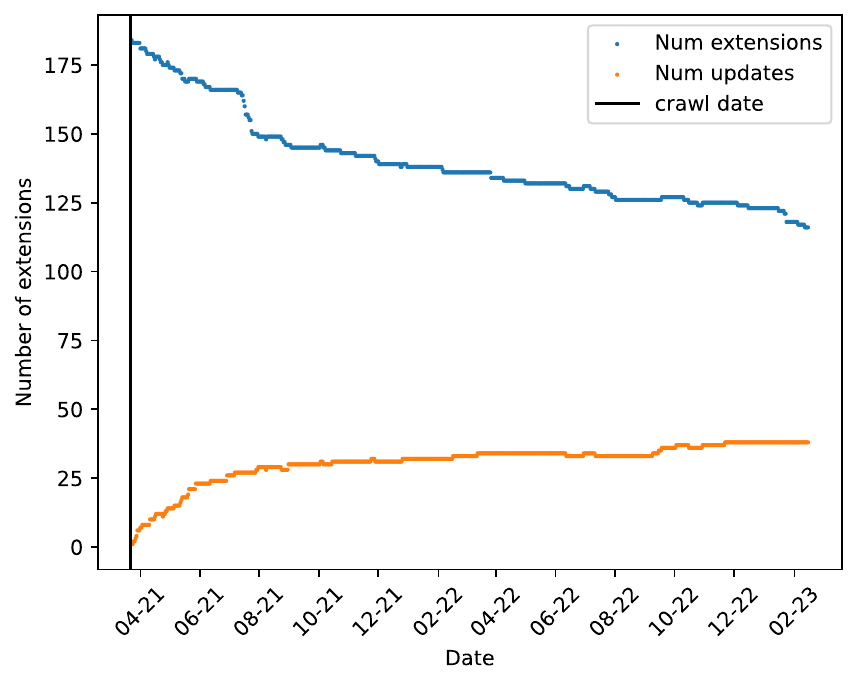}
    \caption{Life cycle of vulnerable extensions%
    \textnormal{--The blue curve represents the vulnerable extensions detected by \textsc{DoubleX}~\cite{fass2021doublex} (crawl: March 16, 2021) that are still in the CWS. In orange, we see the number of those extensions that have received at least one update. The blue minus orange line thus represents a lower bound of the number of extensions that are still in the CWS and vulnerable}}
    \label{fig:DoubleX_remain}
\end{figure}

As shown in \Cref{subsec: updates}, the vast majority of extensions are unmaintained. This is particularly problematic for vulnerable extensions, whose user base can stay at risk \textit{for years} if the vulnerabilities are not fixed.
In fact, in 2021, Fass et al.\@ showed that vulnerable extensions persist in the CWS: of the 193 vulnerable Chrome extensions they found in 2020, 160 (83\%!) were still in the CWS and still vulnerable one year later~\cite{fass2021doublex}.
We extended their experiment from 2021 to 2023. To this end, we contacted the authors and received the list of the 184 vulnerable extensions (including the corresponding versions) they found to be vulnerable in 2021.
In \Cref{fig:DoubleX_remain}, we represent the life cycle of those vulnerable extensions; the blue curve represents the vulnerable extensions detected by their tool, \textsc{DoubleX}, (crawl: March 16, 2021) that are still in the CWS, over time: 116\,/\,184 extensions (63\%) remain in the CWS as of February 2023.
In orange, we see the number of those extensions that have received at least one update.
The blue line minus the orange line thus represents a lower bound of the number of extensions that are still in the CWS and vulnerable (in fact, an update does not necessarily equate to fixing a vulnerability, so our results are a \textit{lower bound} of the number of extensions still vulnerable today).
Overall, we find that at least 78 extensions (over 42\%) are still vulnerable two years after disclosure, impacting over 450k active users.

\summary{Takeaway}{At least 78\,/\,184 extensions (42\%) are still in the CWS and still vulnerable 2 years after disclosure. This shows that, while detecting vulnerable extensions is critical, we also need better incentives to encourage and support developers to \textit{fix vulnerabilities} after disclosure.}

\subsection{Vulnerable JavaScript Library} \label{subsec:vuln-lib}

After \textit{vulnerable extensions}, we finally investigate the prevalence of extensions using \textit{vulnerable JavaScript libraries}.
To detect known vulnerable JavaScript library versions, we leverage RetireJS~\cite{retireJS,retireJSRef} to scan each Chrome extension.
We find that 39,093 extensions (31.5\%) use at least one JavaScript library with a known vulnerability. Those extensions belong to 32,144 distinct developers.
Across all extensions, we identify 87 unique JavaScript libraries with a known vulnerability. We detect over 80k uses of those vulnerable JavaScript libraries, impacting almost 500 million users.
We list the 10 most commonly used vulnerable libraries in \Cref{table:topRetireJs}, along with the corresponding vulnerabilities, and the number of impacted extensions.
The top vulnerable JavaScript library is jQuery, with over 40k extensions loading a vulnerable version (this represents a third of the extensions in the CWS!). As a comparison, the remaining vulnerable libraries we detected are used by less than 10k extensions.
We list the top 10 JavaScript library versions with vulnerabilities in \Cref{table:topRetireJsVers} in the Appendix; the majority of which are jQuery versions.
However, using a vulnerable JavaScript library does not equate to a vulnerable extension, since the vulnerability may not be exploitable from the context of a web application. While some vulnerabilities may be exploitable through Web Accessible Resources, we leave a study of the exploitability of vulnerable libraries for future work.

\begin{table}[t]
\centering
\footnotesize
\begin{tabular}{l|lr}
\toprule

Library & Vulnerability summary & \# extensions      \\

\midrule
jquery & XSS, 3rd party CORS & 41,290 \\
jquery-ui & XSS & 9,632 \\
bootstrap & XSS & 6,746 \\
moment.js & regular expression DoS, path traversal  & 5,408 \\
angularjs & XSS, CSP bypass, DoS & 4,728 \\
jquery-ui-dialog & XSS & 2,223 \\
underscore.js & arbitrary code injection & 1,693 \\
vue & XSS & 1,486 \\
YUI & XSS & 1,027 \\
jquery-validation & regular expression DoS & 960 \\

\bottomrule \end{tabular}
\quad 

\caption{10 most commonly used JavaScript libraries containing known vulnerabilities and used by extensions
	}
\label{table:topRetireJs}
\end{table}

\summary{Takeaway}{Almost a third of extensions (40k) use a JavaScript library with a known vulnerability: we detect over 80k uses of vulnerable libraries, impacting almost 500M extension users.}

\begin{figure}[t]
    \centering
    \includegraphics[width=0.75\columnwidth]{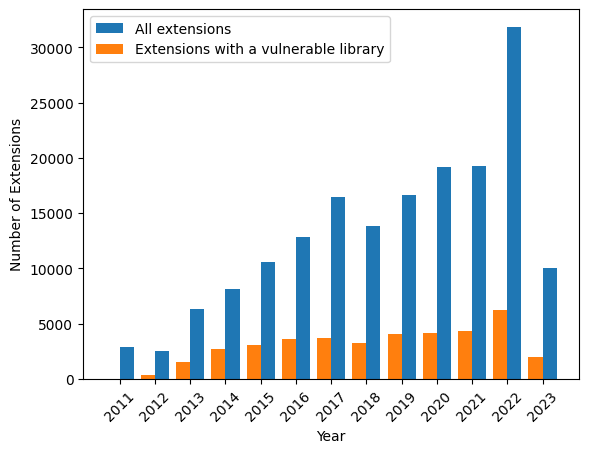} 
    \caption{Number of extensions using a vulnerable library (orange), grouped by last update year, compared to the total number of extensions (blue), by last update year}
    \label{fig:retire_js_updates}
\end{figure}

Next, we investigate how often extensions with a vulnerable library are updated.
In \Cref{fig:retire_js_updates}, we represent the year of the most recent update for each extension in blue and plot the number of extensions that use a library with a known vulnerability, grouped by last update year, in orange.
Surprisingly, we observe that extensions which were updated recently are not less likely to have a vulnerable library than extensions that were updated several years ago. 
Specifically, for extensions with their most recent updates in 2017--2023, between 19.54\% and 24.62\% are using vulnerable libraries.
While extensions with their most recent updates between 2013 and 2016 have a greater fraction of extensions with vulnerable libraries (between 24.49\% to 32.88\%), extensions with their most recent updates in 2011 and 2012 have a smaller fraction (0.81--14.11\%).

Interestingly, and perhaps surprisingly, we find that even when developers are updating their extensions, they are not necessarily maintaining the JavaScript libraries within their extensions. For instance, 21\% of the extensions updated in 2022 and later (8,236 extensions) have a known vulnerable library.
Overall, this provides further evidence that the use of vulnerable libraries persists even when patched library versions are available and even when developers are maintaining their extensions.

\summary{Takeaway}{Even when developers update their extensions, they often \textit{do not} update vulnerable libraries within their extensions, even after a patched version of a vulnerable library has been~released.
We encourage developers to not only maintain their extension source code but also libraries within their extension package.
}

\section{Discussion}

In this section, we first discuss some limitations of our approach.
Then, we summarize our takeaways and recommendations.

\subsection{Limitations}

Since Google does not archive browser extensions that used to be in the CWS, we are dependent on ChromeStats to access longitudinal data. As mentioned in \Cref{subsec:ext-collection}, for some extensions, we only had access to their metadata and not source code.

Another limitation of our paper relates to our code similarity approach. We chose to compute similarity based on the raw source code (\Cref{subsubsec:similar-code-methodo}). However, this approach is not resistant against variable renaming or if a developer chooses to merge scripts in a different order. In the future, we could consider other methods, such as relying on Abstract Syntax Trees~\cite{fass2019hns}.
Regarding our suggestion to use code similarity analysis to identify SNE (\Cref{subsubsec: susp-clusters}), we realize that sharing an identical background or content script with a SNE does not necessarily imply that an extension is also security-critical.
For example, in the case of policy-violating extensions, we would need to also analyze the reported privacy policies; for vulnerable extensions, we would need to review the extension's permissions.
This is why we emphasize this as a tool for flagging extensions for future investigation, not the end-all determination.

Regarding security implications of a lack of maintenance in the CWS, we acknowledge in \Cref{subsec: updates} that poorly maintained extensions do not necessarily lead to security issues: an extension may have achieved its intended functionality when first added to the CWS and not necessitate any further updates.
As for extensions using vulnerable libraries, we emphasize in \Cref{subsec:vuln-lib} that a vulnerability does not equate to exploitability. We leave a study of the exploitability of vulnerable libraries in extensions for future work.

\subsection{Recommendations and Future Work}

We summarize here our main guidelines to guide future research and pave the way for a more secure CWS.

\pseudoparagraph{Volatility in the CWS and Reproducibility}
We observe unexpected volatility of extensions in the CWS. It is important to keep in mind that extensions are frequently being removed from the CWS and new extensions are being added. Additionally, extensions have extremely short life cycles, and extension user counts are unstable.
Given these constant changes, researchers should try to perform their analyses over multiple points in time (e.g., by using data from ChromeStats~\cite{chromestats}) and evaluate the reproducibility of their findings before making claims about broad patterns in the CWS. Specifically, longitudinal studies provide the best assurance of temporal representativeness.
To this end, we encourage researchers to open source their extension sets to ease future work and comparisons.

\pseudoparagraph{SNE Detection and Extension Vetting}
We show that SNE stay in the CWS \textit{for years}.
First, additional efforts are needed to detect such extensions \textit{prior} to their acceptance to the CWS.
For example, existing tools~\cite{fass2021doublex, Yu2023} could help Google engineers flag potential vulnerable extensions.
To detect malware-containing and privacy-violating extensions, we strongly recommend thoroughly scrutinizing extensions submitted to the CWS by developers having published SNE in the past.
Similarly, we encourage researchers and Google engineers to look for extensions with a similar code base to known SNE. Flagged extensions should be more closely monitored, e.g., run through more computationally and time-expensive analyses to detect and reject verified SNE from the CWS.
Second, we suggest to vet extension updates given that ``benign'' extensions can easily turn malicious after permission changes.
Third, we encourage researchers to investigate \textit{why} vulnerable extensions stay in the CWS for years, even after disclosure.
While Google engineers may need to be more active and restrictive when vetting extensions, this also raises the question of who should be responsible for having SNE in the CWS.
For instance, developers may not have any incentives to fix their vulnerable extensions once they have been accepted to the CWS.
In this case, the impacted users should be notified of the security and privacy issues they may be subject to.
This is all the more important as users have a limited understanding of security and privacy issues extensions can induce~\cite{AnkitKariryaa2021}.

\pseudoparagraph{Extension Maintenance}
We obviously encourage developers to maintain their extension source code, including migrating to manifest V3, when applicable.
We also recommend updating the libraries within their extension package.
We leave an investigation of the exploitability of vulnerable libraries for future work.

\section{Related Work}

Prior work focused on detecting \textit{malicious}, \textit{vulnerable}, and \textit{fingerprintable} extensions. Our work is in an orthogonal direction. We investigate what is in the CWS and highlight security implications that were overlooked by prior work.

\pseudoparagraph{Malicious Extensions}
First, malicious extensions are designed by malicious actors to \textit{harm victims}.
For example, some extensions may tamper with security headers~\cite{Agarwal2022}, steal users' credentials, track users~\cite{Weissbacher2017}, spy on them~\cite{Aggarwal2018}, spread malware~\cite{Xing2015}, leak sensitive information~\cite{Chen2018, Starov2017-2},
or have inconsistencies in their privacy practices~\cite{buidetection}.
To defend against these issues, several approaches have been proposed to detect malicious extensions, e.g., by monitoring their behavior~\cite{Wang2012firefoxextensions, Kapravelos2014}, tracking the reputation of developers~\cite{Jagpal2015chromeextensions}, or detecting anomalous ratings~\cite{Pantelaios2020}.
In this paper, we highlight the fact that malicious extensions have different patterns (e.g., number of days in the CWS, user counts, or developers) compared to benign extensions. Such patterns could be leveraged to flag suspicious extensions for additional analyses.

\pseudoparagraph{Vulnerable Extensions}
Second, vulnerable extensions are designed by well-intentioned developers but \textit{contain vulnerabilities}, leading to security or privacy issues.
Prior work focused on vulnerabilities in XPCOM Firefox extensions (XPCOM is now deprecated), based on static information flow tracking~\cite{Bandhakavi2010} and JavaScript namespace vulnerabilities~\cite{Salih2016}.
To detect vulnerable Chrome extensions, some approaches primarily rely on manual analysis~\cite{Carlini2012, Some2019}, others focus on a formal approach~\cite{Calzavara2015}, abstract interpretation~\cite{Yu2023}, or data flow analysis~\cite{fass2021doublex}.
In this paper, we show that vulnerable extensions have patterns quite similar to benign extensions', which makes them more challenging to detect. Besides detection, it is of utmost importance to encourage developers to fix known vulnerabilities that currently pervade the CWS.

\pseudoparagraph{Fingerprintable Extensions}
Third, fingerprintable extensions enable an attacker to \textit{track users} across websites, deanonymize them, or infer sensitive information about them (e.g., related to~health or religion)~\cite{Karami2020}.
Extensions can be fingerprinted by analyzing~their observable side effects.
An attacker can detect the extensions a user has installed by leveraging DOM changes~\cite{Starov2017, Starov2019, Solomos2022-2}, style changes~\cite{Laperdrix2021}, Web Accessible Resources~\cite{Sjosten2017, Karami2020}, user actions~\cite{Solomos2022}, or timing-channels~\cite{SanchezRola2017}.
Several defenses to extension-enumeration attacks have been proposed, such as controlling extensions loaded on a website~\cite{Sjosten2019}, randomizing browser extension fingerprints~\cite{Trickel2019}, or creating a parallel DOM~\cite{Karami2022}.
We leave a study of fingerprintable extensions and a comparison with SNE for future work.

\pseudoparagraph{Untrusted Libraries and Third Parties}
In this paper, we show that extensions are not maintained and use known vulnerable libraries. This is comparable to the npm ecosystem where lack of maintenance causes packages to depend on vulnerable code~\cite{Zimmermann2019}. In npm, one proposed mitigation was to reduce the attack surface by removing unused code~\cite{Koishybayev2020}. We could imagine a similar approach for browser extensions.
More generally, similar issues are observed on the Web (and still an open challenge), when developers do not maintain libraries, allowing their site to be compromised~\cite{Lauinger2017ThouSN}.

\section{Conclusion}

This paper provides a holistic view of the browser extensions’ landscape, within the CWS.
We leverage historical data provided by ChromeStats to broadly investigate and understand what is in the CWS, along with security implications.
Our research \mbox{fundamentally} underlines (a) the extremely \textit{short life cycles of extensions}, (b) the pervasiveness of what we call \textit{``Security-Noteworthy Extensions'' (SNE)}, (c) the presence of \textit{clusters of similar extensions}, and (d) a critical \textit{lack of maintenance in the CWS}.
We are confident that our results can serve as a foundation for researchers and practitioners looking to study and improve the security of the browser extension~ecosystem.

\begin{acks}

We would like to thank ChromeStats~\cite{chromestats}, and more specifically Hao Nguyen, for providing us with Chrome extensions, their metadata, and additional information upon request.
We also thank the anonymous reviewers for their constructive reviews and helpful suggestions.
 Finally, we thank Giovanni Apruzzese and the members of the Stanford Empirical Security Research Group for valuable conversations and feedback.
\end{acks}

\bibliographystyle{ACM-Reference-Format}

\appendix
\section{Appendix}

\begin{table}[hb]
\centering
\resizebox{\columnwidth}{!}{
\begin{tabular}{l l l l l l l}
\toprule

Rank   &  Benign & Malware & Policy violation & Vulnerable  \\
\midrule
1 & storage & storage & storage & tabs\\
2 & tabs & topSites & tabs & storage\\
3 & activeTab & activeTab & cookies & webRequest\\
4 & contextMenus & webRequest & activeTab & webRequestBlocking\\
5 & notifications & webRequestBlocking & management & contextMenus\\
6 & webRequest & cookies & topSites & cookies\\
7 & scripting & management & webNavigation & activeTab\\
8 & unlimitedStorage & webNavigation & webRequest & notifications\\
9 & cookies & tabs & webRequestBlocking & bookmarks\\
10 & webRequestBlocking & unlimitedStorage & notifications & downloads\\

\bottomrule 
\end{tabular}
}
\quad

\caption{Top 10 API-based permissions for benign and SNE}
\label{table:permissions_top}
\end{table}

\begin{table}[!bh]
\centering

\footnotesize
\begin{tabular}{l l r}
\toprule

Rank & URL  &     Occurrences \\

\midrule
1 &   http://opensource.org/licenses/mit-license.php & 426,224 \\
2 &           http://apache.org/licenses/LICENSE-2.0 & 383,040 \\
3 &                              http://jqueryui.com &  99,984 \\
4 &      http://opensource.org/licenses/BSD-3-Clause &  87,712 \\
5 &               http://code.google.com/p/crypto-js &  65,376 \\
6 &  http://code.google.com/p/crypto-js/wiki/License &  65,248 \\
7 &               http://opensource.org/licenses/MIT &  65,056 \\
8 &                        http://jquery.org/license &  59,552 \\
9 &                    http://codemirror.net/LICENSE &  58,240 \\
10 & https://github.com/twbs/bootstrap/blob/master/LICENSE &  57,312 \\

\bottomrule 
\end{tabular}
\quad 

\caption{Most popular URLs and number of occurrences of a URL across all JavaScript files of all extensions
\textnormal{(total of over 50k URLs)}
}
\label{table:esprima_url}
\end{table}

\begin{table}[b]
\centering

\footnotesize
\begin{tabular}{l l r r}
\toprule

Rank & Second-level domain  &  \# unique URLs  & \# unique extensions \\

\midrule
1 &            github.com & 9,685 & 62,285 \\
2 & ecma-international.org & 516 & 9,976 \\
3 &             github.io & 1,128 & 8,461 \\
4 &            whatwg.org & 737 & 7,491 \\
5 &     stackoverflow.com & 3,036 & 6,793 \\
6 &           chrome.com & 698 & 6,422 \\
7 &                w3.org & 870 & 6,413 \\
8 &            google.com & 1,297 & 6,319 \\
9 &           mozilla.org & 1,388 & 6,088 \\
10 &        opensource.org & 48 & 5,030 \\
\bottomrule \end{tabular}
\quad 

\caption{Most popular second-level domains with the number of unique extensions they appear in
\textnormal{(total of 18,551 second-level domains)}}
\label{table:esprima_rootdomain_unique}
\end{table}

\begin{table}[!b]
\centering

\footnotesize
\begin{tabular}{l|lr}
\toprule

Library & Version       & \# extensions \\

\midrule

jquery &	3.3.1 &	4,744 \\
jquery &	3.4.1 &	3,700 \\
jquery &	3.2.1 &	3,326 \\
jquery-ui &	1.12.1 &	2,242 \\
jquery &	3.1.1 &	2,144 \\
jquery-ui &	1.11.4 &	2,015 \\
jquery &	2.1.1 &	1,970 \\
jquery &	2.1.4 &	1,921 \\
jquery &	2.0.0 &	1,871 \\
jquery &	1.10.2 &	1,808 \\

\bottomrule \end{tabular}
\quad 

	\caption{Top 10 JavaScript library versions containing known vulnerabilities and used by extensions
    }
\label{table:topRetireJsVers}
\end{table}

\begin{figure}[bh]
    \centering
    \includegraphics[scale=0.465]{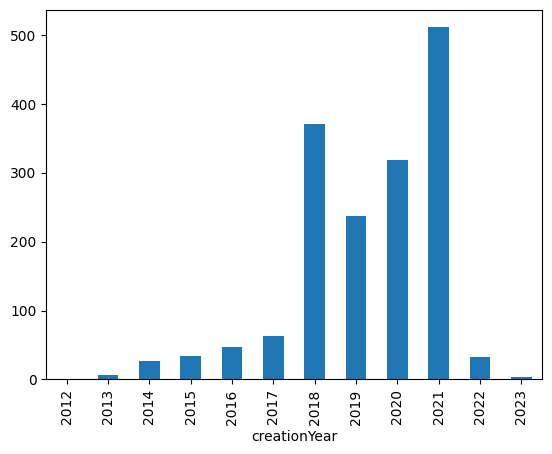}
    \caption{Number of extensions created each year using Extensionizr~\cite{extensionizr}
    }
    \label{fig:extensionizr}
\end{figure}

\end{document}